\documentclass[aps,pra,reprint,groupedaddress]{revtex4-1}
\usepackage{graphicx}
\usepackage{color}
\graphicspath{{figure/}}
\usepackage{amsmath}
\usepackage{hyperref}
\graphicspath{{figure/}}
\begin{document}

\title{Miniaturized optical system for a chip-based cold-atom inertial sensor}

\author{S. Hello$^{1,2}$, H. Snijders$^3$, B. Wirtschafter$^3$, A. Boutin$^4$, L. Fulop$^4$, F. Seguineau$^1$, C. I. Westbrook$^2$, A. Brignon$^3$, M. Dupont-Nivet$^3$}

\affiliation{${}^{1}$Thales AVS France SAS, 40 rue de la Brelandière, 86100 Châtellerault, France \\
${}^{2}$Laboratoire Charles Fabry, Institut d'Optique Graduate School, 2 av. Augustin Fresnel, 91127 Palaiseau, France \\
${}^{3}$Thales Research and Technology France, 1 av. Augustin Fresnel, 91767 Palaiseau, France \\
${}^{4}$Exail, 11 av. De Canteranne, 33600 Pessac, France}

\date{\today}

\begin{abstract}
We miniaturized the complex optical system responsible for the cooling, pumping and imaging of an on-chip based cold atom inertial sensor. 
This optical bench uses bonded miniature optics and includes all the necessary optical functions.
The bench has a volume of 35x25x5~cm$^3$. 
We developed a laser frequency lock adapted to the optical bench using saturated absorption in a rubidium cell. 
The entire laser source based on frequency doubling of 1.56~$\mu$m fiber lasers, including the control system and the saturated absorption module, fits in a $5U$-rack. Using the miniaturized bench, we realized two and three dimensional magneto optical traps for Rubidium 87 atoms.
\end{abstract}

\maketitle


\section{Introduction}

Cold atom interferometers arouse great interest for the realization of inertial sensors due to their excellent stability and sensitivity \cite{Geiger2020}. 
Combining three accelerometers, three gyrometers and one clock, one can build an inertial measurement unit, which is an essential part of a navigation system, embedded on several kind of carriers such as aircraft, submarine, drone and satellite \cite{Barbour2010}. 
This device, when fixed on moving platforms, delivers information which enables to compute inertial navigation solutions \cite{Jekeli2005,Savage1998a,Savage1998b}. 
This technique permits the positioning of a moving platform without any external information. 
Knowing the initial position, using a map and the Newton’s equation of motion, one can find the current position. 
This constitutes an autonomous alternative to Global Navigation Satellite Systems (GNSS) hybridized inertial systems, which are subject to jamming and spoofing.  

However, conventional cold atom inertial sensors based on free fall atoms remain too large for many inertial navigation applications. 
In order to overcome this limitation, on-chip atom interferometers enable a drastic size reduction as well as decoupling the sensitivity from the size of the sensor \cite{Bohi2009,Schumm2005,DupontNivet2014}. 
An atom chip \cite{Reichel1999} allows trapping atoms close to its surface thanks to on-chip DC currents. 
Moreover, using microwave waveguides on chip, we can coherently manipulate the trapped atoms \cite{Treutlein2004,Bohi2009}. 
We are developing an atom chip inertial sensor with a chip that allows the trapping and the manipulation of a thermal cloud (with a temperature a few times above the bose-einstein condensation threshold) near its surface throughout the whole duration of the inertial measurement \cite{Ammar2014}.

To use cold Rubidium atom technologies for atomic clocks in space \cite{Aveline2020,Laurent2020}, for on board gravimetry \cite{Bidel2018,Bidel2020}, or microgravity experiments \cite{Geiger2011,Muntinga2013,Rudolph2015,Becker2018} a lot of effort has been conducted to develop and ruggedized the laser system needed to cool and manipulate atoms.
As an example, some mobile gravimeters use laser diodes in extended cavities at 780~nm with semiconductor amplifiers \cite{Schmidt2011,Schmidt2011b, Cheinet2006,Merlet2014,Zhang2018} and with a free space optical system for routing and switching all the laser beams needed in the device. This approach has been ruggedized to operate in space \cite{Laurent2006, Leveque2015,Elliott2018}, a drop tower and a sounding rocket \cite{Schiemangk2015,Schkolnik2016,Dinkelaker2017,Pahl2019, Strangfeld2021} where a volume of 43 liters has been achieved \cite{Pahl2019}.
Another approach is to use frequency doubling of telecom lasers at 1.56~$\mu$m \cite{Lienhart2007,Carraz2009, Theron2015,Theron2017, Diboune2017,Dingjan2006, Nyman2006, Stern2009, Menoret2011}. This allows higher power at 780~nm \cite{Sane2012} than with direct emission at 780~nm using diode lasers and semiconductor amplifiers and the whole system can be fully fiber-based \cite{Leveque2014,Legg2017,Diboune2017}.
A non-exhaustive review of these systems is presented in table \ref{Tab_Comp}.

\begin{table*}
\caption{\label{Tab_Comp} Example of laser systems to cool and manipulate Rubidium atoms}
\begin{ruledtabular}
  \begin{tabular}{ p{1.5cm} || p{2.5cm} | p{4cm} | p{9.0cm} } 
 Reference & Volume & Included functions & Comments \\ 
 \hline\hline
 \cite{Pahl2019} & 43~L & cooling and Raman/Bragg lasers for ${}^{87}$Rb and ${}^{41}$K & withstand 43~g in operation (drop tower and sounding rocket) in microgravity experiments, free space system with laser diodes and semiconductor amplifiers \\ 
 \hline
 \cite{Diboune2017} & - & cooling and Raman lasers for ${}^{85}$Rb, ${}^{87}$Rb and ${}^{133}$Cs & fully fiber-based system with frequency doubling for Rb and frequency sum for  Cs, operated in a gravimeter \\
 \hline
 \cite{Menoret2011} & fit in a 19 inches rack structure & cooling and Raman beams for ${}^{87}$Rb and ${}^{39}$K & frequency doubling for Rb and K fully fiber-based except the doubling modules that are free space, operated in microgravity experiments during parabolic flight on a plane \\
 \hline
 \cite{Merlet2014} & - & cooling and Raman beams for ${}^{87}$Rb & free space system with two laser diodes (cooling and pumping frequency) injecting the same semiconductor amplifier operated in a gravimeter \\
 \hline
 \cite{Zhang2018} & 45x45x16 cm$^3$ $\approx$ 32 L & cooling and Raman lasers for ${}^{85}$Rb & free space system with laser diodes and semiconductor amplifiers, used to operate a mobile gravimeter \\ 
 \hline
 \cite{Leveque2015} & 53x33x20 cm$^3$ $\approx$ 35 L & cooling and selection of Cs & space qualified, use four laser diodes and four others for redundancy, operated in a microwave atomic clock  \\
 \hline
 \cite{Elliott2018} & - & cooling an Bragg beams for ${}^{87}$Rb, ${}^{39}$K and ${}^{41}$K & space qualified, use laser diodes and semicondictor amplifiers \\
 \hline
 This work & 42 L = 4.4 L + 37.6 L & cooling (3D MOT and 2D+ MOT), optical pumping and detection for ${}^{87}$Rb & Raman lasers not needed for interferometer \cite{Ammar2014}, 35x25x5 cm $\approx$ 4.4 L (beams routing and switching) and $5U$-rack $\approx$ 37.6 L (fiber lasers, amplifiers, fiber doubling and lock system)  \\ 
 \end{tabular}
\end{ruledtabular}
\end{table*}

In this paper, we present our recent efforts implemented to reduce the volume of the optical system needed for our on chip interferometer \cite{Ammar2014} which is similar to the one used for on chip atomic clocks \cite{DupontNivet2025}. 
Using the proprietary bonding technology for miniature optics developed by Exail, we realized a routing and switching module for the cooling of Rubidium 87 in a volume of 35x25x5~cm$^3\approx 4.4$~L.  
This module includes all the optical functions needed for the preparation of the cooling, optical pumping and detection beams.
To inject this module, we develop a laser source based on fiber lasers at 1.56~$\mu$m, amplified with fiber amplifiers and frequency doubled using all-fibered Periodically Poled Lithium Niobate (PPLN) waveguide which fits in the volume of a $5U$-rack $\approx 37.6$~L.
To characterize those two modules, we realize a three Dimensional Magneto Optical Trap (3DMOT) loaded with a two Dimensional Magneto Optical Trap (2DMOT). 
In the following, we present the bench structure (Section \ref{Section_2}), the laser system adapted to it (Section \ref{Section_3}) and the obtained results (Section \ref{Section_4}).
 
\section{The miniaturized optical system}
\label{Section_2}

\subsection{Optical system for a chip based cold atom inertial sensor}

We are realizing a chip based inertial sensor using ultracold 87 Rubidium atoms. 
For the realization of our interferometer, we need to cool, trap and prepare the atoms in a specific state \cite{Huet2013,DupontNivet2016,Wirtschafter2022}.
The first cooling step is Doppler cooling which is combined with a magnetic field to create a Magneto-Optical Trap (MOT). 
For the realization of the MOTs, we need to prepare several laser beams at specific frequencies. 
The current free space optical system used for the atom cooling in our lab has a surface area of 20~000~cm$^2$ and a height of 15~cm.
Using this optical system, the atom preparation and the interferometry sequence described in \cite{Wirtschafter2022}, we obtained a Ramsey interferometer with a state-selective spatial splitting \cite{Wirtschafter2022,Bohi2010}.
However, this optical system stays too voluminous to be embedded in moving platforms, thus we have to reduce its size.

\subsection{The miniaturized optical system}

\begin{figure}
\centering  \includegraphics[width=0.48\textwidth]{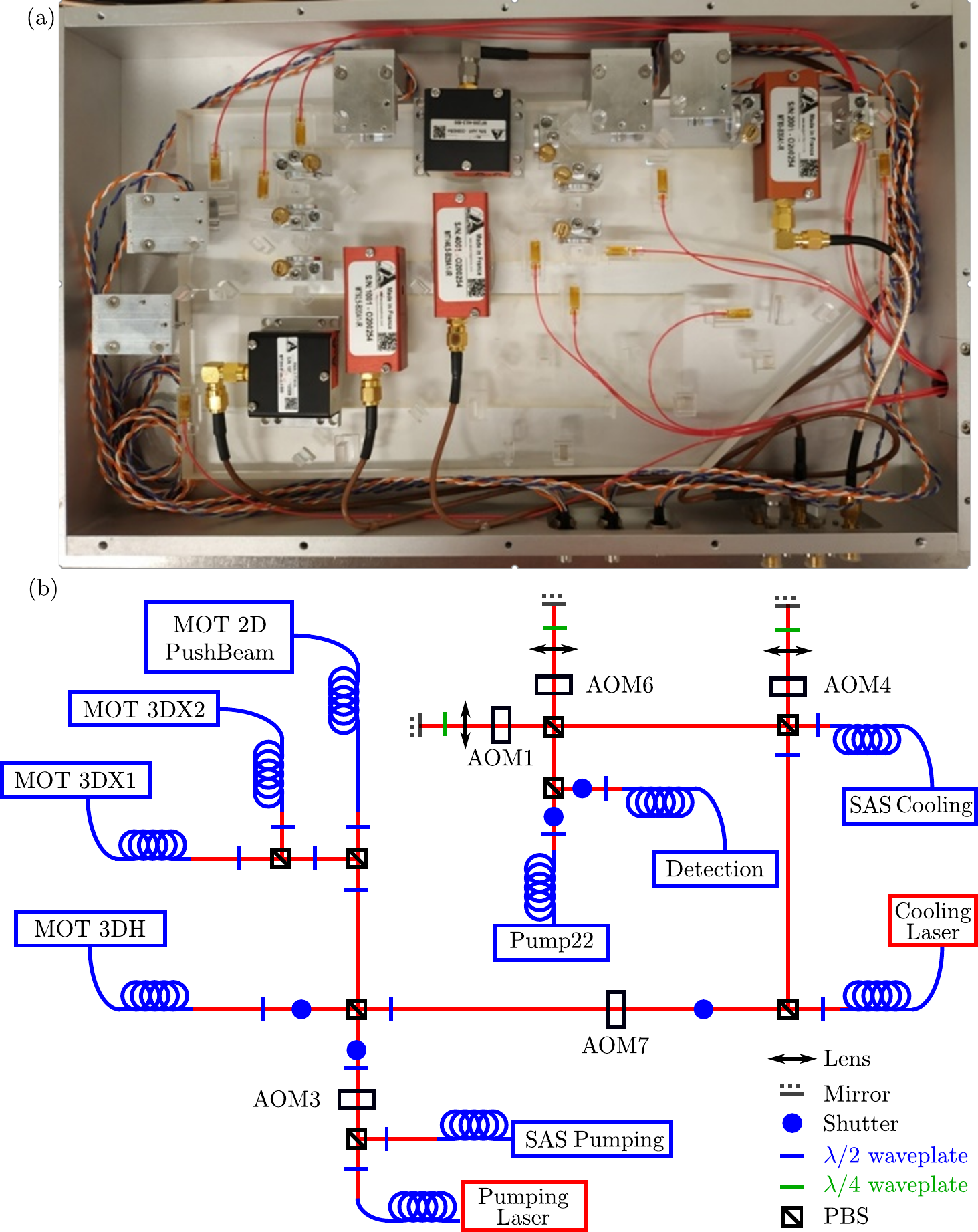}
\caption{\label{fig_01} (Color online) (a)  Miniaturized optical bench with a volume of 35x25x5~cm$^3$. 
(b) Functional diagram of the optical system implemented for to prepare the laser beams for atom cooling, pumping and detection. 
Red (respectively blue) rectangles are for the inputs (respectively outputs).}
\end{figure}

The reduced optical bench is realized with bonded miniature optics as shown in Figure \ref{fig_01}.a. 
It prepares all the laser beams for the experiment. 
Figure \ref{fig_01}.b shows a complete optical diagram of the bench. 
The bench has two fiber inputs in red and eight fiber outputs in blue. 
The three outputs MOT~3DH, MOT~3DX1 and MOT~3DX2 are the three beams used for the 3DMOT (see figure \ref{fig_08} for explaination of the beam labels). 
The output MOT~2D/PushBeam is used for a pre-cooling in two dimensions by realizing a 2DMOT \cite{Dieckmann1998,Schoser2002}.
The PushBeam moves the pre-cooled atoms from the 2DMOT to the 3DMOT. 
In this paper, the other outputs are not used. 
SAS Pumping (respectively SAS Cooling) allows sending a fraction of the pumping (respectively cooling) laser power to lock its frequency for example on using a Saturated Absorption Spectroscopy (SAS). 
Pump22 is used for optical pumping of atoms in state $\left|F=2,m_F=2\right>$.
Detection is used to detect atoms at the outputs of the interferometer.

\begin{figure}
\centering  \includegraphics[width=0.40\textwidth]{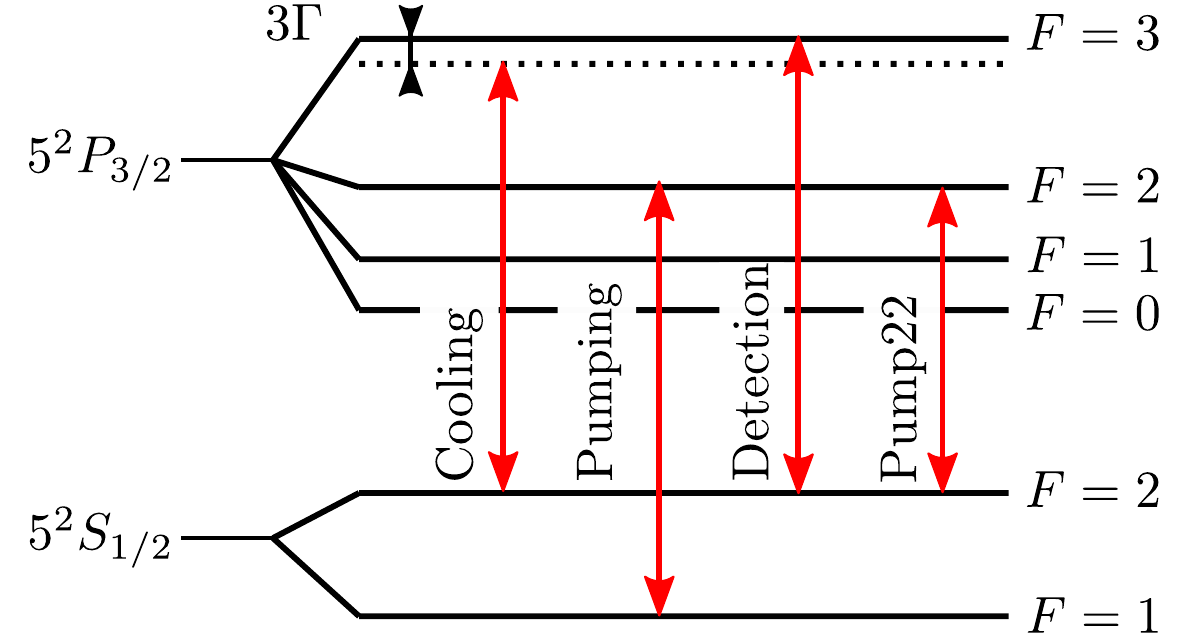}
\caption{\label{fig_02} (Color online) Hyperfine structure of Rubidium 87 atoms - D2 line \cite{Steck2003}. 
Cooling and pumping transitions are used for the atom cooling. 
$3\Gamma$ is the detuning of the cooling laser during the MOTs. $\Gamma = 6$~MHz is the transition natural width. Detection beam is tuned to $\left|5^2 S_{1/2},F=2\right> \rightarrow \left|5^2P_{3/2},F=3\right>$ to detect atoms at the output of the interferometer. 
Pump22 beam is tuned to $\left|5^2 S_{1/2},F=2\right> \rightarrow \left|5^2P_{3/2},F=2\right>$ to pump atoms in $\left|F=2,m_F=2\right>$.}
\end{figure}

The frequency used for the cooling corresponds to the transition $\left|5^2 S_{1/2},F=2\right> \rightarrow \left|5^2P_{3/2},F=3\right>$ of the hyperfine structure of Rubidium 87 for D2 line as shown in Figure \ref{fig_02}. 
The cooling of an atom works on a closed cycling transition.
However, this one is not closed because the frequency difference between $\left|5^2P_{3/2},F=3\right>$ and $\left|5^2P_{3/2},F=2\right>$ is only about 266~MHz, thus some atoms can go to $\left|5^2P_{3/2},F=2\right>$ leading to a possible decay in $\left|5^2 S_{1/2},F=1\right>$. 
After a large number of absorption-emission cycles, all atoms will fall into that state and there will be no longer atoms in the cooling cycle.
Therefore, we need a second laser, the pumping laser that is resonant with the $\left|5^2 S_{1/2},F=1\right> \rightarrow \left|5^2 P_{3/2},F=2\right>$ transition, to recycle atoms fallen into the $\left|5^2 S_{1/2},F=1\right>$ state back into the cooling cycle. The two inputs of the bench are for the two cooling and pumping lasers, together cooling the atoms. 
The cooling laser is tuned to the cooling transition and the pumping laser is tuned to the pumping transition (see Figure \ref{fig_02}). 

Inside the optical bench, Polarization Beam Splitters (PBS) allows for separation and recombination of the beams. 
Each output needed for the realization of the 2DMOT and 3DMOT (MOT~3DX1, MOT~3DX2, MOT~3DH, MOT~2D/PushBeam) has the cooling and pumping frequencies. 
Acousto-Optical Modulators (AOM) in the bench allow for adjusting the laser frequencies. 
We can shift the frequency of the cooling beam during the optical molasses (using AOM4 which shifts the cooling frequency before sending it to the saturated absorption spectroscopy), the optical pumping (using AOM6) and atom detection (using AOM1). 
For those shifts, we use AOMs in a double-pass configuration \cite{Donley2005}. 
The pumping laser beam is locked on the transition between $\left|5^2 S_{1/2},F=1\right>$ and the crossover between $\left|5^2 P_{3/2},F=1\right>$ and $\left|5^2 P_{3/2},F=2\right>$ using saturated absorption spectroscopy, which is not the pumping frequency, thus the AOM3 in the bench shifts the laser to the pumping frequency.
Adjustable half wave plates let us adjust the optical powers on each output. 
Mechanical shutters extinguish beams during the interferometry sequence and assure that no light arrives to the atoms.

\section{Atom cooling using the miniaturized optical system}
\label{Section_3}

For the realization of the MOTs, we need to set up lasers and lock systems adapted to the optical bench. 
Once lasers are stabilized and tuned to the cooling and pumping frequencies, they can be injected into the bench. 

\subsection{Laser system}

\begin{figure}
\centering  \includegraphics[width=0.48\textwidth]{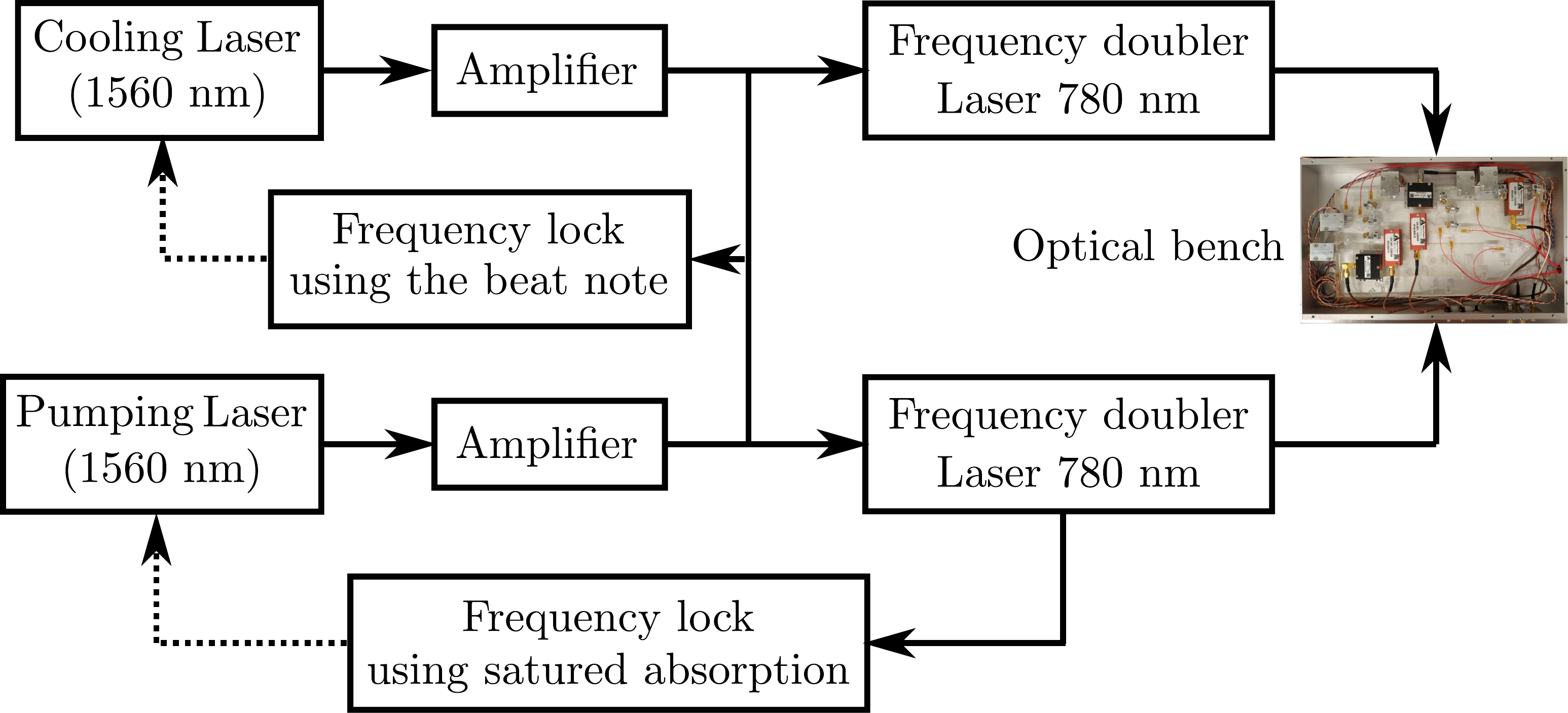}
\caption{\label{fig_03} (Color online) Laser system adapted to the miniaturized optical system of the bench. 
The two 1560~nm lasers are amplified, and frequency doubled to generate 780~nm laser beams. 
We lock the pumping laser using saturated absorption spectroscopy and the cooling laser using the beat note with the first laser. 
Solid lines are optical paths and dashed lines are electrical signal paths.}
\end{figure}

Figure \ref{fig_03} shows a schematic of our laser system. 
We use two telecom fiber lasers at 1560~nm. Each laser delivers a power of 12~mW. 
We amplify the laser beams with miniature telecom optical fiber amplifiers to have an optical power of 840~mW for the cooling and 440~mW for the pumping laser. 
They are then frequency doubled by a PPLN waveguide to reach 780~nm leading to the input powers, in the optical bench, of 380~mW for the cooling laser and 110~mW for the pumping laser.

We lock the pumping frequency using a saturated absorption spectroscopy. 
We lock the cooling frequency using the beat note with the pumping laser. 
We sent 2~mW of the pumping laser at 780~nm to the spectroscopy. 
We use 3~mW of the cooling laser and 3~mW of the pumping beam at 1560~nm to generate the beat note.

\subsection{Pumping laser frequency lock}

\begin{figure}
\centering  \includegraphics[width=0.48\textwidth]{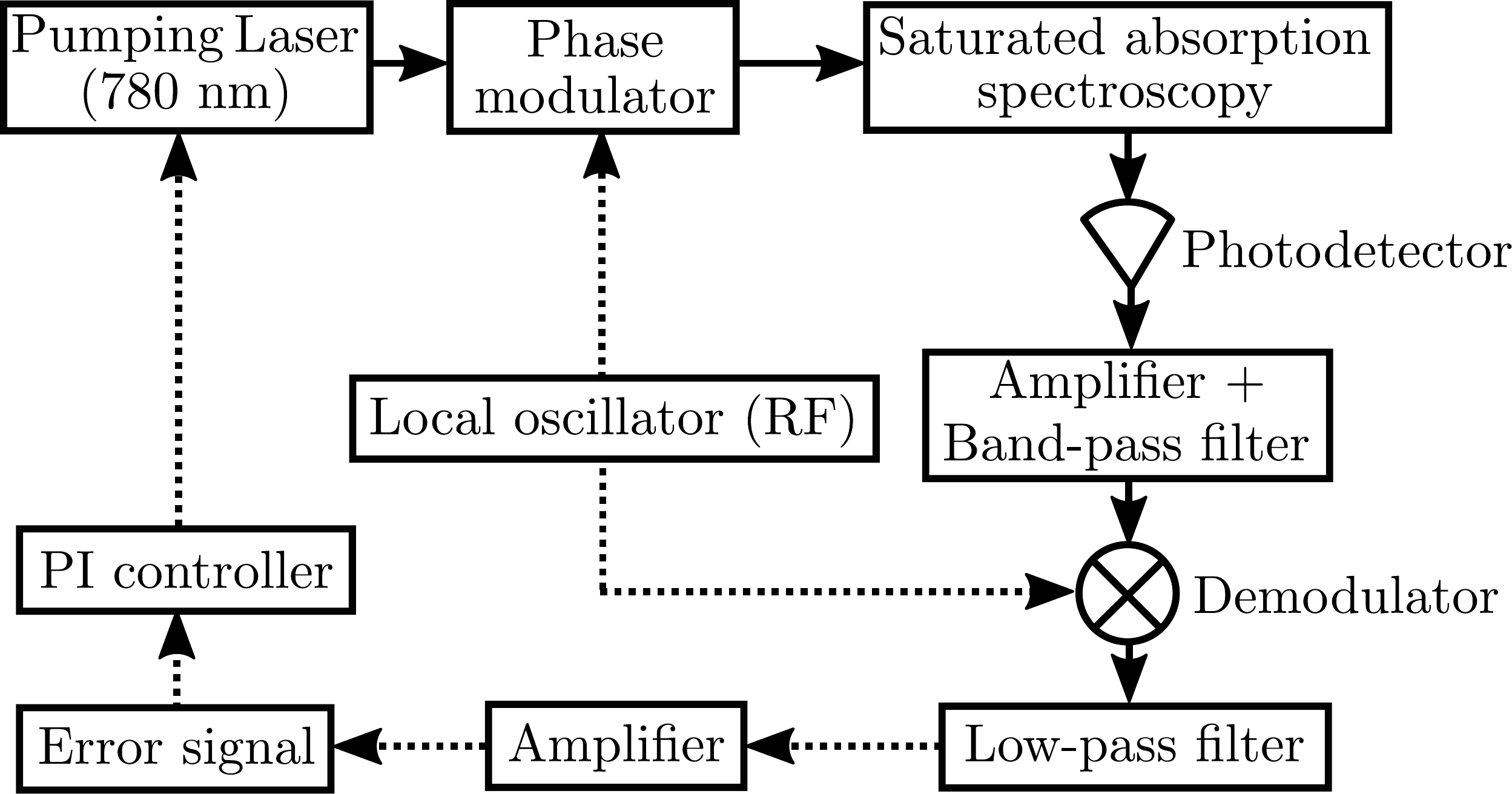}
\caption{\label{fig_04} (Color online) Experimental layout at 780~nm for frequency lock of the pumping laser. 
The laser is locked using the Pound-Drever-Hall scheme on a frequency reference provided by saturated absorption spectroscopy. 
Solid lines are optical paths and dashed lines are electrical signal paths. 
The electrical signal going to the pumping laser controls its frequency.}
\end{figure}

To lock the pumping laser, we use the Pound-Drever-Hall scheme \cite{Black1998,Black2001} and saturated absorption spectroscopy. 
The saturated absorption provides a strong stable frequency reference that enables to bring out the hyperfine structure of atoms. 
Figure \ref{fig_04} illustrates the experimental layout adopted.

Before sending 2~mW of the pumping (at 780~nm) to the saturated absorption sepectrocopy, we modulate its frequency with a fiber phase modulator driven around 620~kHz. 
A photodetector recovers the saturated absorption signal which is then amplified with a carrefully designed electronic and filtered with a band-pass filter between 400~kHz and 900~kHz. 
Then, the signal is demodulated using a SBDIP demodulator by multiplying this signal with the sign of signal used to drive the phase modulator. 
The demodulated signal passes through a low-pass filter of 5~kHz and is amplified to create the error signal.

\begin{figure}
\centering  \includegraphics[width=0.48\textwidth]{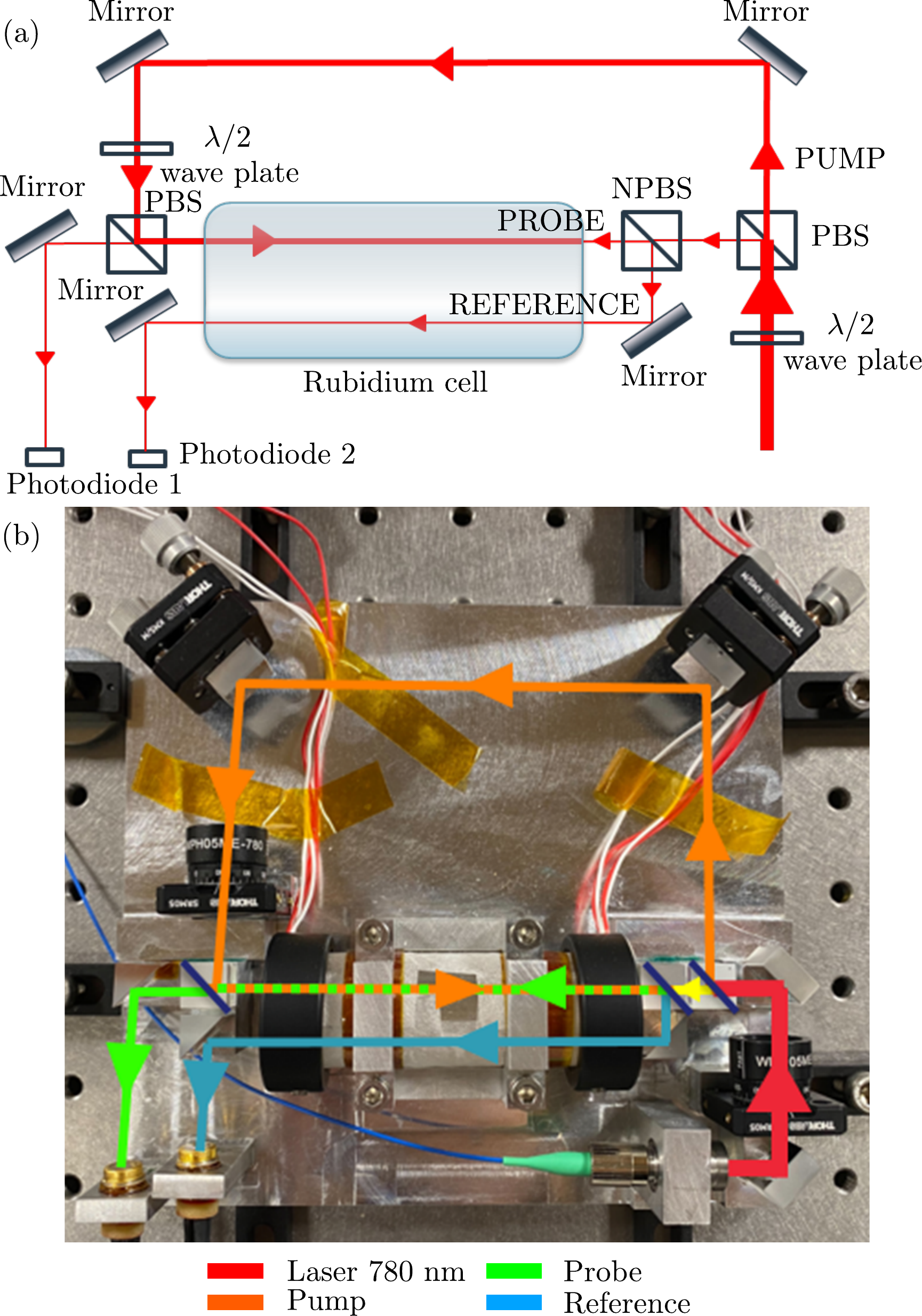}
\caption{\label{fig_05} (Color online) (a) Schematic of the saturated absorption spectroscopy experiment used to lock the pumping laser. 
(b) Picture of the experimental saturated absorption spectroscopy. 
The size of the set-up is about 15x15~cm$^2$. }
\end{figure}

Figure \ref{fig_05}.a shows the saturated absorption set up made with bounded miniature optics.
The footprint is about 15x15~cm$^2$.
The 780~nm pumping laser beam passes through a first half-wave plate as well as a first PBS adjusted such that the pump beam is about 1~mW and the probe beam is about 100~$\mu$W. 
A 50:50 non-polarizing beam splitter separates the probe beam in two identical parallel beams that are sent through the Rubidium cell: the probe beam and the reference beam. 
The pump beam is reflected on two mirrors and returns in opposite direction (see Figure \ref{fig_05}) such that it overlaps the probe beam in a counter propagating way in the Rubidium cell. 
Finally, probe and reference beams finish their course on photodiodes. 
The photodiode 1 signal gives the absorption spectrum expanded by Doppler effect with the hyperfine structure highlighted. 
The photodiode 2 signal gives the absorption spectrum expanded by Doppler effect. 
By subtracting both signal, we extract the hyperfine structure of the rubidium. 
For the pumping laser lock, we are interested in the transition between $\left|5^2 S_{1/2},F=1\right>$ and $\left|5^2 P_{3/2},F\right>$ states of the Rubidium 87 structure.

\begin{figure}
\centering  \includegraphics[width=0.35\textwidth]{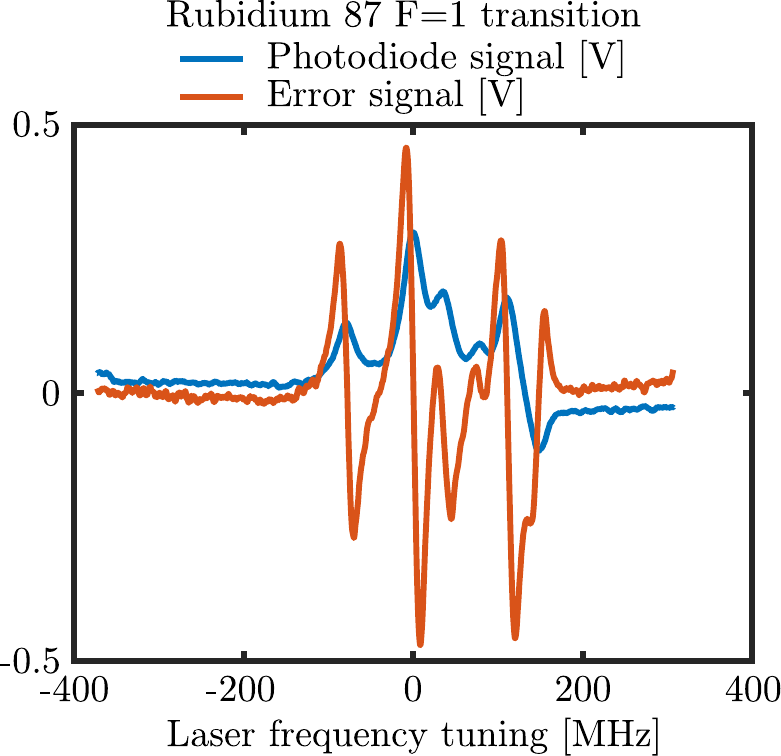}
\caption{\label{fig_06} (Color online) 
Saturated absorption signal is in blue (this signal is 30 times amplified for better reading) for the $|5^2 S_{1/2},F=1>$ state of ${}^{87}$Rb atoms.
Error signal is in orange.
The zero corresponds to 384.234~604~760~THz. }
\end{figure}

Figure \ref{fig_06} shows in blue the obtained saturated absorption signal, and in orange, its derivative genrated by the described electronic. 
We lock the pumping laser frequency to the most intense transition which is the cross over between $\left|5^2 P_{3/2},F=1\right>$ and $\left|5^2 P_{1/2},F=2\right>$. 
Hence, we set up a proportional integrator servo controller. 
The signal is fed back to control the piezoelectric wedge inside the fiber laser. 
The long term drift of the laser in lock condition is about 600~kHz over 5 hours. 
It is mainly limited by instabilities in some offset signals in the control electronics.

\subsection{Cooling laser frequency lock}

\begin{figure}
\centering  \includegraphics[width=0.48\textwidth]{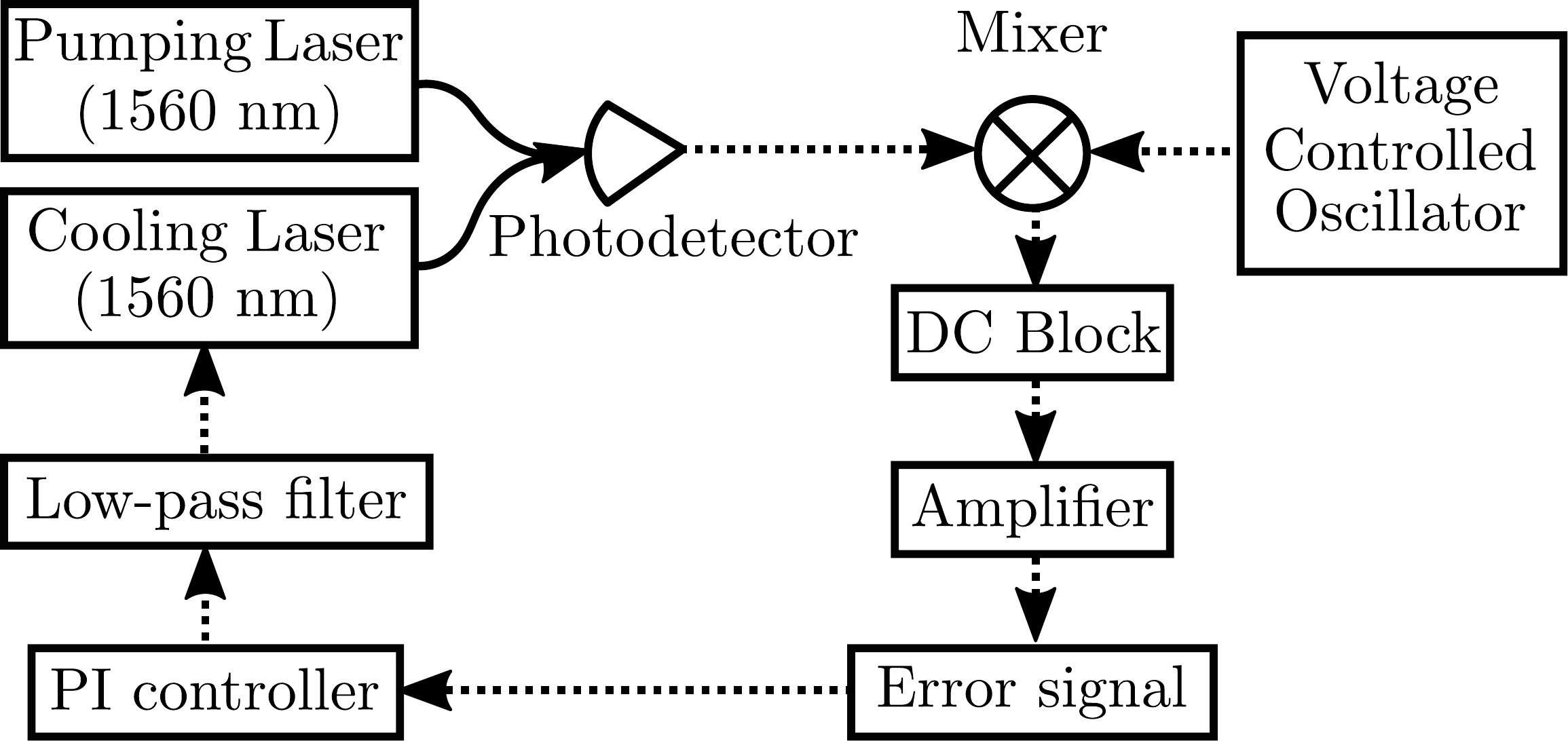}
\caption{\label{fig_07} (Color online) Experimental layout for 1560~nm laser frequency lock using the beat note with a locked laser. 
The pumping laser is locked on a rubidium transition. 
The cooling laser is locked using the beat note with the first. 
Solid lines are optical paths and dashed lines are electrical signal paths. 
The signal going to the cooling laser controls its frequency. }
\end{figure}

The cooling laser is locked using the beat note with the pumping, as illustrated in Figure \ref{fig_07}. 
Taking into account the frequency shift due to the AOMs, the 1560~nm beat frequency must be around $f_{bn}\approx$~3.145~GHz. 
This beat note is sent to a high-speed fiber-coupled detector and then multiplied with the signal of a Voltage Controlled Oscillator (VCO) oscillating at $f_{vco}$.
We keep only the frequency difference component $f_{bn}-f_{osc}$ and adapt the VCO frequency in order to have a signal around $f_{bn}-f_{osc}\approx$~2.1~MHz. 
This signal is sent through an electronic generating a voltage proportional to the frequency $f_{bn}-f_{osc}$ which is fed to a proportional integrator servo controller.
The controler output is fed back to control the piezoelectric wedge inside the fiber laser.
Changing the voltage on the VCO allows changing the cooling frequency.
 
The entire control system of the two lasers fits in a rack of 45x40x12~cm$^3$ and promises to reduce further the volume of the sensor.

\section{2DMOT and 3DMOT realized with the miniaturized optical system}
\label{Section_4}

Once the laser beams are locked to the two frequencies, we can introduce them to the inputs of the miniaturized optical bench and adjust the half wave plates to reach the following power level on the outputs: on MOT~3DX1 and MOT~3DX2 there is about 28~mW of cooling and 13~mW of the pumping, and on MOT~3DH there is about 28~mW of cooling laser and 6~mW of the pumping laser.
On MOT~2D/PushBeam, there is about 85~mW of cooling laser and 17~mW of pumping laser. 
There is a few mW on the other outputs. 
We recover on the outputs 190~mW of cooling laser and 50~mW of pumping laser. 
Therefore, the miniaturized optical bench has a transmission of about 50\%.

\begin{figure}
\centering  \includegraphics[width=0.38\textwidth]{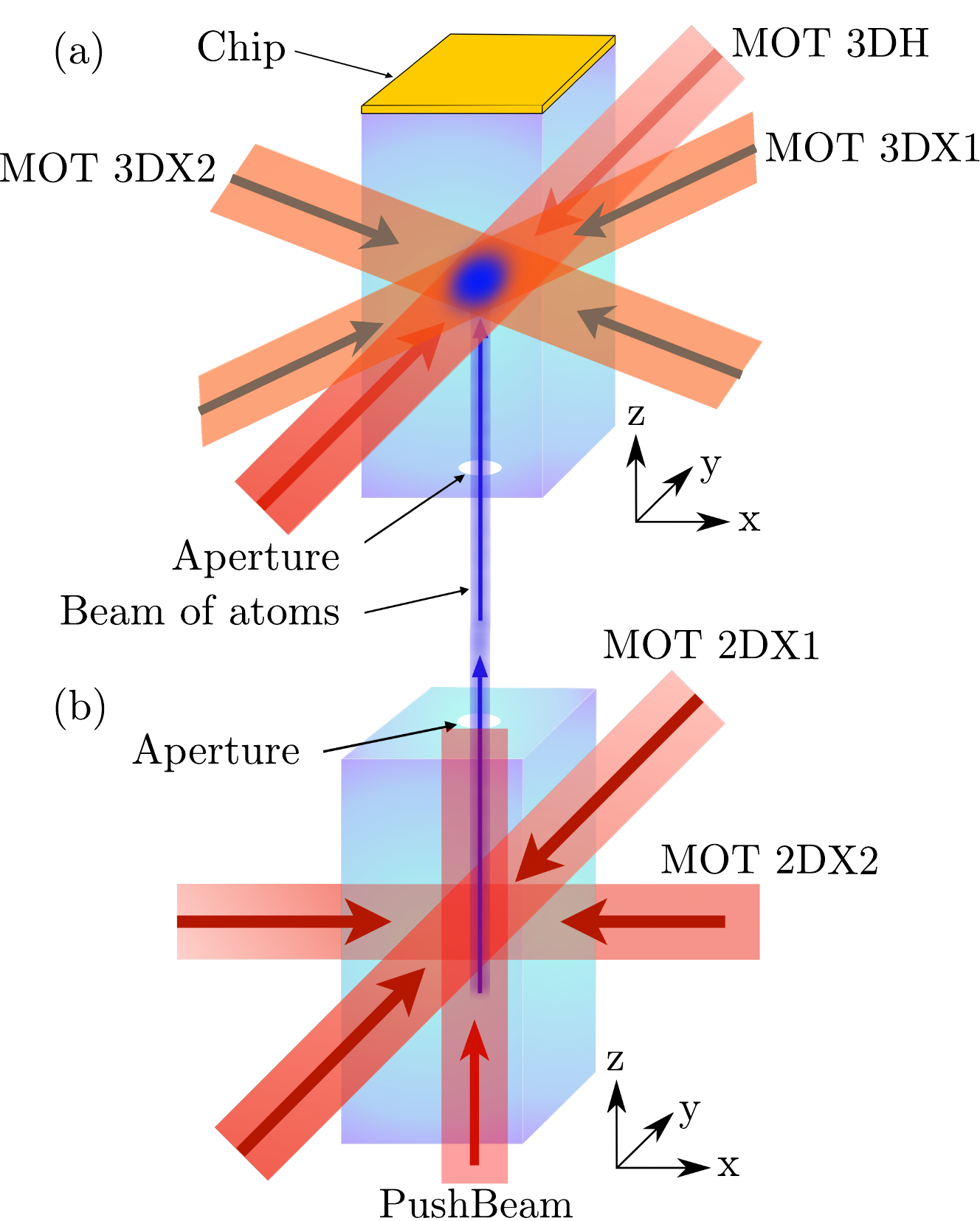}
\caption{\label{fig_08} (Color online) (a) Scheme of the 3DMOT obtained with three pairs of counter propagating beams and coils (not shown). 
MOT~3DH beam is oriented along the y axis and the MOT~3DX1 and MOT~3DX2 beams are orthogonal and in the xz plan. The blue cloud is the 3DMOT. 
(b) Scheme of the 2DMOT obtained with two pairs of counter propagating beams and magnets (not shown). MOT~2DX1 beam is along y, MOT~2DX2 along x, and the PushBeam is along z.
This results in a collimated atom jet (in blue) that charges the upper chamber.
If needed the PushBeam (along the z axis) allows to increase the atom transfer rate from the 2DMOT to the 3DMOT. }
\end{figure}

To demonstrate the cooling with those lasers and miniature bench, we realize a 2DMOT and a 3DMOT. 
Figure \ref{fig_08} shows the layout of the vacuum cell on which we realize this demonstation. 
References \cite{Metcalf2012,Cohen2011} describe in detail the physical principle of a magneto-optical trap.

\begin{figure*}
\centering  \includegraphics[width=0.78\textwidth]{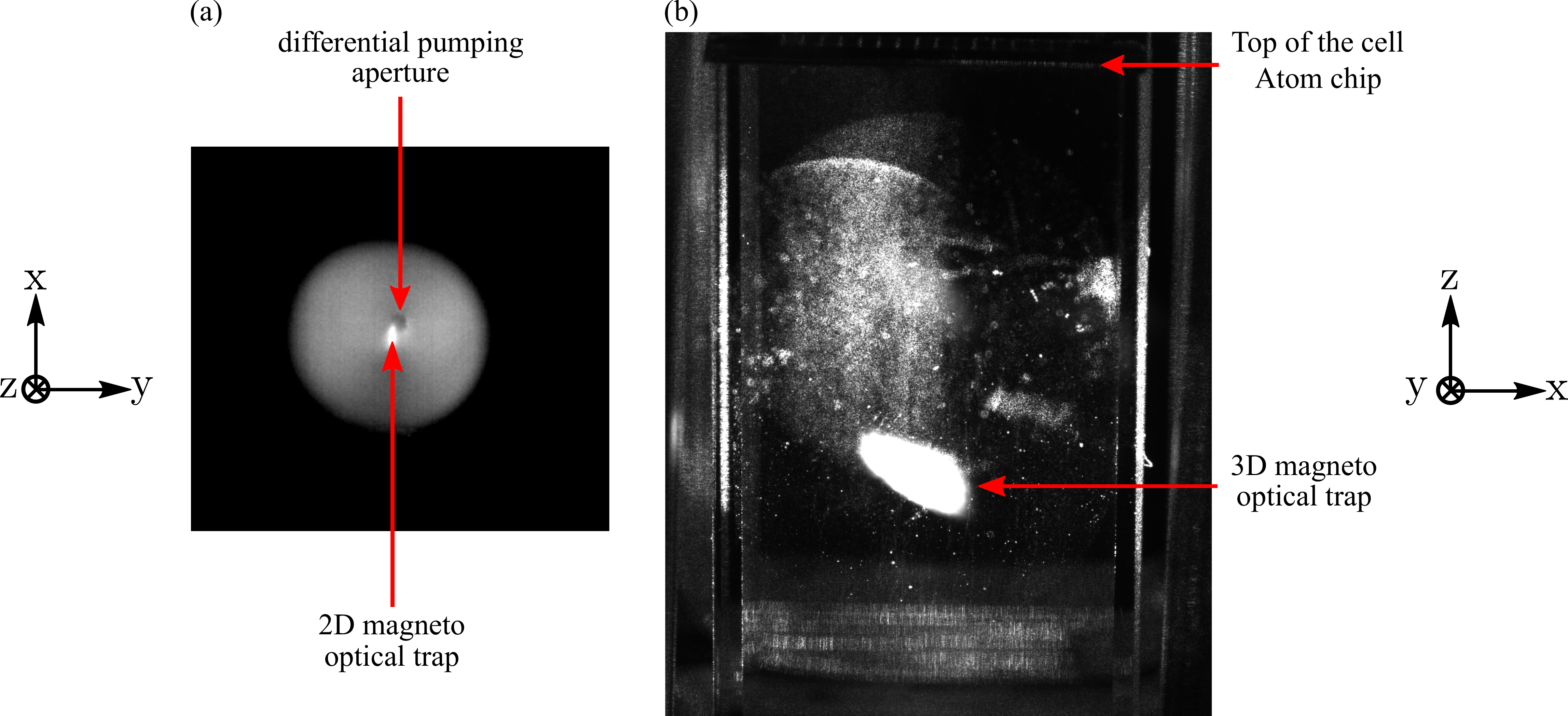}
\caption{\label{fig_09} (Color online) (a) 2DMOT and (b) 3DMOT obtained using the beams prepared by the miniaturized optical bench.  
By fluorescence, we see the atoms in the 2DMOT and 3DMOT. 
Atom number in the 3DMOT is about $2.5\cdot 10^8$. Axis are the same than figure \ref{fig_08}.}
\end{figure*} 

The vacuum cell of our sensors is made up of two chambers as described in \cite{DupontNivet2016,Wirtschafter2022}, the first one for the 2DMOT, the second one for the 3DMOT. 
Both are connected through a differential pumping aperture of 750~$\mu$m diameter.
With two pairs of counter propagating laser beams, the PushBeam and a magnetic field gradient created by magnets, we realize a 2DMOT to cool atoms in the xy plane in the first chamber (see Figure \ref{fig_08}.b). 
The magnets are designed to create a magnetic field gradient in the plane xy  and invariant along the z axis with a zero in the center of the cell where the atoms will accumulate. 
We obtain a thin beam of atoms which velocity direction is oriented along the z axis. 
The PushBeam pushes the atoms from the first chamber to the second chamber along the z axis through the differential pumping aperture. 
Then, with three pairs of counter propagating laser beams and a magnetic field gradient along the x, y and z axis created by two coils in anti-Helmoltz configuration, we realize a 3DMOT in the second chamber \cite{Gajda1994,Devlin2016} (see Figure \ref{fig_08}.a).  
The realization of a 2DMOT before a 3DMOT allows charging rapidly a larger number of atoms in the 3DMOT, while keeping the vacuum below $10^{-10}$~mbar in the second chamber\cite{Farkas2010}. 
In the 3DMOT and 2DMOT, the counter propagating laser beams  are created with mirrors on which the beam reflect and overlap with the original beam.

We first connect to the cell the output MOT~2D/PushBeam of the optical bench to create a 2DMOT. 
To create the two beams of the 2DMOT a 50:50 polarization maintaining fiber splitter is added at the ouput MOT~2D/PushBeam. 
For the whole test reported in this article, we did not sent light in the PushBeam. 
The 2DMOT is displayed in Figure \ref{fig_09}.a.
Then, the 2MOT is used to load a 3DMOT.
We take the three outputs MOT~3DX1, MOT~3DX2 and MOT~3DH from the optical bench and insert them to the upper cell. 
With one pair of coils whom axis is parallel to the MOT~3DH beam, we create a magnetic field with a gradient of 4~G.cm$^{-1}$ with a zero at the intersection of the three beams of the 3DMOT. 
We obtain the 3DMOT shown in Figure \ref{fig_09}.b. 
We image the top of the upper chamber with the atom chip that closes the top. 
The 3DMOT is approximately 600~$\mu$m by 250~$\mu$m and contains $2.5\cdot 10^8$~atoms. Atom number is measured with the fluorescence of the 3DMOT \cite{Lewandowski2003}. 

\section{Conclusion}

The miniaturized optical bench presented in this paper brings together all the optical functions necessary for the cooling, pumping and imaging. 
We shrink a 200x200x15~cm$^3$ optical system down to a 35x25x5~cm$^3$ one. 
The laser beams used to characterize the optical system were frequency locked using a saturated absorption spectroscopy in a Pound-Drever-Hall scheme on the pumping laser and using the beat note between the cooling and pumping lasers.
We succeed to cool atoms using the miniaturized optical system by realizing a 3DMOT which is loaded by a 2DMOT. 
The obtained atom number ($2.5\cdot 10^8$) is comparable to the one obtained with the 200x200x15~cm$^3$ optical system \cite{DupontNivet2025} using the same vacuum cell and is compatible with other measurement on vacuum cell with similar layout \cite{Farkas2010}. This 3DMOT atom number is high enough to cool few $10^4$ atoms via evaporative cooling in an on chip magnetic trap down to few 100~nK \cite{DupontNivet2025, Farkas2010}. Those numbers allow reaching on atom chip clock with a $10^{-13}$ relative stability \cite{DupontNivet2025, Szmuk2015} and would allow on-chip acceleration sensitivity of 1~$\mu$g \cite{DupontNivet2014}.
The demonstrated volume reduction of the optical system paves the way to a chip based sensor for inertial navigation of moving platforms.
 
Further developments will be to improve the electronic offset stability in the pumping lock system to achieve better long term performances of the laser frequency lock. 
Moreover, we can improve the 3DMOT loading and atom number by setting up the PushBeam and fine tuning the optical alignement of the MOT beams.

\vspace*{0.3cm}
\begin{acknowledgments}
This work has been carried out within the NIARCOS project ANR-18-ASMA-0007-02 funded by the Agence de l'Innovation de Défense (AID) in the frame of its 2018 Astrid Maturation programs.
\end{acknowledgments}
 


\bibliography{biblio}

\end{document}